\documentclass[aps,prd,twocolumn,showpacs,superscriptaddress,nofootinbib,preprintnumbers]{revtex4-2}
\usepackage{amsmath}
\usepackage{tikz}
\usepackage{tikz-feynman}
\tikzfeynmanset{
  compat=1.1.0
}
\tikzset{every path/.append style={thick}}
\usepackage{amsfonts}
\usepackage{amssymb}
\usepackage{enumerate}
\usepackage{color}
\usepackage{graphicx}
\usepackage{bm}
\usepackage[colorlinks=true,citecolor=blue,urlcolor=blue]{hyperref}
\usepackage{afterpage}
\usepackage{multirow}
\usepackage{appendix}
\usepackage[table]{xcolor}
\usepackage{feynmp-auto}
\definecolor{tabblue}{RGB}{31, 119, 180}
\definecolor{darkblue}{RGB}{0, 0, 120}
\definecolor{tabred}{RGB}{214, 39, 40}
\definecolor{tabgreen}{RGB}{44, 160, 44}
\definecolor{tabgray}{RGB}{100, 100, 100}
\usepackage{float}
\usepackage{placeins}
\def\beq{\begin{equation}}
\def\eeq{\end{equation}}
\def\bea{\begin{eqnarray}}
\def\eea{\end{eqnarray}}

\newcommand{\mpl}{M_{\rm pl}}

\def\lsim{\mathrel{\rlap{\lower4pt\hbox{\hskip1pt$\sim$}}
     \raise1pt\hbox{$<$}}} 
\def\gsim{\mathrel{\rlap{\lower4pt\hbox{\hskip1pt$\sim$}}
     \raise1pt\hbox{$>$}}}

\begin{document}
\preprint{\hbox{NORDITA-2025-038, UTWI-19-2025}}
\title{Thermal Gravitons from Warm Inflation}

\author{Gabriele Montefalcone}
\email{montefalcone@utexas.edu}
\affiliation{Texas Center for Cosmology and Astroparticle Physics, Weinberg Institute for Theoretical Physics, Department of Physics, The University of Texas at Austin, Austin, TX 78712, USA}
\author{Barmak Shams Es Haghi}
\email{shams@austin.utexas.edu}
\affiliation{Texas Center for Cosmology and Astroparticle Physics, Weinberg Institute for Theoretical Physics, Department of Physics, The University of Texas at Austin, Austin, TX 78712, USA}
\author{Tao Xu}
\email{tao.xu@ou.edu}
\affiliation{Homer L. Dodge Department of Physics and Astronomy, University of Oklahoma, Norman, OK 73019, USA}
\author{Katherine Freese}
\email{ktfreese@utexas.edu}
\affiliation{Texas Center for Cosmology and Astroparticle Physics, Weinberg Institute for Theoretical Physics, Department of Physics, The University of Texas at Austin, Austin, TX 78712, USA}
\affiliation{The Oskar Klein Centre, Department of Physics, Stockholm University, AlbaNova, SE-10691 Stockholm, Sweden}
\affiliation{Nordic Institute for Theoretical Physics (NORDITA), 106 91 Stockholm, Sweden}

\begin{abstract}
In warm inflation (WI), the persistent thermal bath that is sustained by
 dissipative interactions with the inflaton field produces a stochastic background of gravitational waves (GWs). In this paper we study the production and evolution of these GWs.
Specifically, we investigate the emission of  thermal gravitons (gravitons emitted by a thermal bath) from  scattering with particles in the bath and then the evolution of the corresponding high-frequency GWs. We find that the bulk of thermal graviton production in WI occurs during the transition to radiation domination after inflation. Further, the energy density of thermal gravitons is enhanced by roughly one to two orders of magnitude compared to that in a radiation-dominated scenario with the same reheating temperature. We also calculate the spectrum of the resulting stochastic GW background and find that it has a distinctive shape, consisting of a peak at high frequencies ($\sim 100\,{\rm GHz}$) and an almost flat spectrum extending to low frequencies. The peak arises from emission of sub-horizon modes that follow the temperature of the bath. The flat part of the spectrum corresponds to the modes that exit the horizon during WI and re-enter during radiation domination and reflects the approximately
constant temperature of the thermal bath during WI. We show that the detection prospects for the high-frequency peak of the GW spectrum, while improved slightly compared to the radiation-dominated case, still remain challenging. 
The thermal spectrum’s low-frequency plateau is typically subdominant to the amplitude of the standard vacuum tensor modes from inflation, although WI models can exist where the thermal graviton plateau surpasses the vacuum contribution without exceeding current observational limits on the tensor-to-scalar ratio. Furthermore, we calculate the thermal graviton contribution from WI to dark radiation and show that WI models are generally expected to satisfy current observational bounds including those from the cosmic microwave background (CMB). 

\end{abstract}

\maketitle

\section{Introduction}

In this paper, we explore the emission and evolution of high-energy gravitons, and the corresponding stochastic gravitational wave (GW) background, generated by collisions in the thermal bath present during warm inflation (WI).
In the WI scenario, the potential energy of the inflaton dominates throughout the inflationary
phase, while continuously sourcing a persistent but subdominant radiation bath with nearly constant temperature above the corresponding Hubble expansion rate~\cite{Berera:1995ie,Berera:1995wh,Berera:1998px}. Through these dissipative effects, WI enables a smooth transition of the Universe to a radiation-dominated era without requiring a separate reheating phase. The existence of a thermal bath in WI implies that particle species must interact sufficiently to stay thermalized throughout inflation. Once such a bath is established,  these scattering processes inevitably lead to graviton emission. This provides a generic source of a stochastic GW background with distinct features at both high and low frequency regimes that we explore in detail in this work.
 
\medskip
The fundamental mechanism underlying this graviton production involves the constituent particles of the system experiencing acceleration due to collisions and scattering off one another, with the rate of scattering depending on their number density and interactions. As a result, the system loses energy through the emission of GWs~\cite{Weinberg:1972kfs}. Under ordinary conditions in the current Universe, even a fully thermalized system emits only a small amount of energy into gravitons. For reference, the solar gravitational radiation power is $\sim 10^8\,\mathrm{W}$~\cite{Weinberg:1965nx,2024arXiv240718297G},
far below the $\sim 10^{26}$ W of electromagnetic radiation power~\cite{IAUInter-DivisionA-GWorkingGrouponNominalUnitsforStellarPlanetaryAstronomy:2015fjh}. In the early Universe, however, when the primordial thermal bath could reach much higher temperatures, this emission can be much larger and manifest as a stochastic GW background. There are two regimes for this emission process. In the first regime, the case relevant to this paper, there is sufficient time between collisions to prevent interference, and each collision contributes independently to graviton production. This leads to the microscopic emission of high-energy (or equivalently, high-frequency) gravitons, with typical momenta $k \sim T$, where $T$ is the temperature of the thermal bath. In the second regime, relevant for low-energy graviton emission with $k \ll T$ (not the case in this paper), the system behaves collectively like a fluid. In this regime, graviton production arises from macroscopic hydrodynamic fluctuations ~\cite{Weinberg:1972kfs, Ghiglieri:2015nfa, Ghiglieri:2020mhm},  though this contribution is negligible in the WI context and will not be considered further in this work.

In this paper, we refer to gravitons emitted from a thermal bath as {\it thermal gravitons}. However, this does not imply that the gravitons themselves are thermalized, which would require temperatures near the Planck scale~\cite{Kolb:1990vq}. 

Extensive studies have explored the irreducible stochastic GW background sourced by the thermal bath in the radiation-dominated era, assuming the particle content consists solely of the Standard Model (SM) species~\cite{Ghiglieri:2015nfa,Ghiglieri:2020mhm,Ringwald:2020ist}. Since thermal emission peaks at energies comparable to the temperature, and the frequency of the emitted GWs redshifts in the same way as the temperature, the present-day peak of the GW spectrum is expected to lie in the microwave range; more precisely, around 80\,GHz. While the peak always occurs at the same frequency, its amplitude is sensitive to the highest temperature of the bath ever attained (e.g. the reheating temperature). In principle, the resultant stochastic GW background contributes into dark radiation and can affect the expansion rate of the Universe. Observational constraints on dark radiation arise from the predicted abundances of light elements during Big Bang Nucleosynthesis (BBN), and from the decoupling of the cosmic microwave background (CMB) of photons, which occur when the temperature of the Universe is $\sim 0.1\,\text{MeV}$ and $\sim 0.1\,\text{eV}$, respectively. For reasonable choices of the maximum temperature of the SM thermal bath, the contribution from thermal gravitons remains well below these bounds~\cite{Ringwald:2020ist}. 
The effects of additional degrees of freedom motivated by physics beyond the SM~\cite{Ringwald:2020ist}, hidden sectors~\cite{Drewes:2023oxg}, and non-standard cosmologies~\cite{Muia:2023wru,Bernal:2024jim,Ai:2025fqw}, have also been explored in this context. For instance,  it is shown that, in a radiation-dominated epoch, adding new, weakly-coupled degrees of freedom to the SM does not enhance the resulting GW signal; while these particles increase the emission of thermal gravitons, they also contribute to the Hubble expansion rate, leading to greater redshifting of the signal~\cite{Ringwald:2020ist,Muia:2023wru}.

\medskip

Here we explore the emission, evolution, and spectrum of GWs from microscopic particle collisions in the thermal bath produced during WI. We note that the contribution of hydrodynamic fluctuations to GW emission in the WI context has been previously explored in Refs.~\cite{Qiu:2021ytc,Klose:2022rxh}, where it was shown to exhibit the universal $\sim k^3$ spectral shape with a model-dependent magnitude controlled by the shear viscosity. During the WI inflationary period, thermal gravitons are continually produced but then redshifted away due to the quasi-exponential expansion of the Universe.
Indeed we will show that the production of thermal gravitons by WI peaks during the transition to the radiation domination phase and their energy density is roughly one to two orders of magnitude larger than that of the  GWs emitted during radiation domination with the same reheating temperature. By calculating the spectrum of the associated stochastic GW background, we will show that it has a distinctive shape which consists of a sharp peak at high frequencies, followed by an almost flat part extending to low frequencies. The peak represents the emission of sub-horizon modes that follow the temperature of the bath, while the flat part  originates from the modes that exit the horizon during WI and re-enter during radiation domination. Furthermore, we explore detection prospects of the thermal stochastic GW background produced by WI and find that the high-frequency peak, while enhanced compared to the radiation-dominated case, still remains challenging to detect. We also investigate the conditions on the evolution of the temperature and the Hubble expansion rate during
WI that enable the low-frequency plateau of the thermal spectrum to surpass the vacuum tensor spectrum, while keeping the tensor-to-scalar ratio within current observational limits. Finally, we determine the conditions under which observations can constrain the contribution to dark radiation of thermal gravitons produced by WI.
   
This paper is organized as follows. After a brief review of WI in Section~\ref{sec:WIReview}, we  review the emission of high-frequency GWs from a thermal bath in Section~\ref{sec:thermalgraviton}, and then apply this to the thermal bath in a WI setup to calculate the energy density and the spectrum of the resulting stochastic GW background. In Section~\ref{sec:constraints}, we discuss the constraints and detection prospects for the
thermal gravitons produced during WI. Our conclusions are presented in Section~\ref{sec:conclusions}. In Appendix~\ref{sec:appA}, we provide upper bounds on the spectrum of thermal gravitons produced during WI, as well as their contribution into dark radiation. To make an analytical treatment possible, we perform the calculation for a minimal WI setup in which the inflationary phase is described by single-field slow-roll dynamics with a monomial potential and constant dissipation strength.

\section{Warm Inflation: A Brief Review}
\label{sec:WIReview}
In WI, the inflaton field $\phi$ dissipates its energy into light degrees of freedom at a rate faster than the Hubble expansion rate, $H$. This allows the produced particles to thermalize and form a subdominant yet persistent radiation bath during inflation. The evolution of the homogeneous
inflaton field in the presence of a thermal bath with temperature $T$ is then governed by the following set of equations:
\begin{align}
    &\ddot{\phi}+(3H+\Upsilon)\dot{\phi}+dV(\phi)/d\phi=0,\label{eq:inflaton} \\
    & \dot{\rho}_r+4H\rho_r=\Upsilon \dot{\phi}^2,\label{eq:rhor} \\
    & H^2=\left(\rho_\phi+\rho_r\right)/\left(3 M_{\rm pl}^2\right),\label{eq:Hubble}
\end{align}
where $\rho_\phi=V(\phi)+\dot{\phi}^2/2$ is the inflaton energy density, $\rho_r=(\pi^2/30)g_\star(T)T^4$ is the energy density of the thermal bath with $g_\star(T)$ number of relativistic degrees of freedom, and $M_{\rm pl}\equiv 1/\sqrt{8\pi G}\approx 2.4 \times 10^{18}\,$GeV.  The Hubble expansion rate $H$ is determined by the Friedmann equation. The dissipation term, $\Upsilon(\phi,T)$, characterizes the rate at which the inflaton field's energy is transferred into radiation and its specific form is determined by the details of the underlying microphysical model~\cite{Berera:1998px}. 

For the remainder of this work, we aim to keep our discussion general and model-independent. However, to support our results with a concrete example, we consider a quartic inflaton potential, $V(\phi)=\lambda \phi^4$ along with a linear dissipation rate $\Upsilon\propto T$. This choice is motivated by the Warm Little Inflaton model~\cite{Bastero-Gil:2016qru, Bastero-Gil:2019gao}, which provides the simplest successful realization of WI consistent with CMB observations, as well as by recent developments involving an axion-like inflaton field coupled solely through Standard Model gauge interactions~\cite{Berghaus:2025dqi}. We stress, however, that the key qualitative features of our findings remain independent of any specific model choice.

It is customary to introduce the dimensionless dissipation strength $Q$, which characterizes the efficiency with which the inflaton transfers its energy into radiation:
\begin{equation}
    Q\equiv \Upsilon/(3H).
\end{equation}
We consider two representative scenarios characterized by initial dissipation strengths of $Q_{\rm ini} = 10^{-2}$ and $Q_{\rm ini} = 1$. In Ref.~\cite{Freese:2024ogj}, Eqs.~\eqref{eq:inflaton}, \eqref{eq:rhor}, and \eqref{eq:Hubble} for a quartic inflaton potential are evolved using these initial dissipation strengths to yield 60 e-folds of inflation. The self-interacting coupling $\lambda$ is chosen to reproduce cosmological observables consistent with the CMB constraints set by \textit{Planck}~\cite{Planck:2018jri}. For our analysis, we use the $T(N_e)$ and $H(N_e)$ corresponding to these initial dissipation strengths, which are provided in Ref.~\cite{Freese:2024ogj}, after verifying that the effect of thermal graviton production on the evolution of $T(N_e)$ and $H(N_e)$ is negligible.

\section{Thermal Graviton Production}
\label{sec:thermalgraviton}
In this section, after reviewing the production of thermal gravitons from a thermal bath in a general cosmological setting,  we first compute the resulting energy density and then derive the spectrum of the corresponding stochastic GW background generated from a WI scenario.

\subsection{Energy Density of Gravitational Waves}
The evolution of the energy density of gravitons produced via particle scattering in a thermal bath of temperature $T$ with respect to cosmic time, $t$, is given by~\cite{Ghiglieri:2015nfa,Ghiglieri:2020mhm}:
\bea
\frac{d \rho_{\rm GW}}{d t}+4H\rho_{\rm GW}=\frac{4T^4}{M^2_{\rm Pl}}\int \frac{d^3k}{(2\pi)^3}\hat{\eta}(T,\hat k),
\label{eq:rhoGW_1}
\eea
where the dimensionless source term, $\hat{\eta}(T,\hat{k})$, encodes the production rate and kinematic distribution of gravitons from particle scatterings in the thermal plasma at a temperature $T$, and $\hat{k}\equiv k/T$ where $k$ is the physical momentum. The exact form of $\hat \eta(T, \hat k)$ depends on the underlying microphysics, including the particle content and interaction rates of the thermal bath, and has been studied for a generic quantum field theory coupled to gravity in Refs.~\cite{Ghiglieri:2015nfa,Ghiglieri:2020mhm,Ringwald:2020ist}. The leading-log estimate of $\hat{\eta}(T,\hat{k})$ around its peak can be expressed as~\cite{Ghiglieri:2015nfa,Ringwald:2020ist}:
\begin{equation}
    \hat \eta (T, \hat k)\approx A_{\eta}(T)\frac{\hat k}{e^{\hat k}-1},
    \label{eq:leadlogeta}
\end{equation}
where $A_{\eta}(T)$ captures the dependence on the Debye thermal masses of the bath particles, normalized by the temperature, and is expected to be $\sim \mathcal{O}(1)$. A full leading order calculation of $\hat{\eta}(T,\hat{k})$ can be found in Ref.~\cite{Ghiglieri:2020mhm}, which includes logarithmic terms that depend on $\hat k$. However, without loss of generality, and to maintain a model-independent set-up for our calculations, we use the leading-log estimate of $\hat{\eta}(T,\hat{k})$, i.e., Eq.~\eqref{eq:leadlogeta}, with $A_{\eta}(T_{\rm max})=1$ for the rest of this work. It is worth noting that, as expected, the integrand that governs the graviton production, $\hat k^2 \hat\eta(T,\hat k)$, peaks at a physical momentum of the order of the thermal bath temperature, specifically for $k\simeq 2.82\, T$.

The production of thermal gravitons can be understood as a freeze-in process~\cite{Hall:2009bx,Elahi:2014fsa}, via a non-renormalizable operator suppressed by $M_{\rm pl}$. 
Production of particles from the thermal bath during WI via freeze-in (WIFI) has been studied in detail in Ref.~\cite{Freese:2024ogj}.\footnote{For an application of the WIFI framework in the context of the seesaw mechanism,   see~\cite{deSouza:2024oaz}. Its application to the production of dark matter via gravitational interactions can be found in~\cite{Wang:2025duy}.} Two key features of particle production in WIFI are identified: (i) it is enhanced compared to freeze-in during a radiation-dominated era with the same reheating temperature, and (ii) the bulk of the production always occurs before the onset of the radiation-dominated phase, at a time set by the evolution of the temperature and the Hubble expansion rate within a given WI scenario, as well as the mass dimension of the non-renormalizable operator responsible for the production process. Therefore, we expect the production of thermal gravitons to peak sometime before the onset of radiation domination, with their energy density enhanced compared to the thermal graviton production in a radiation-dominated era with the same reheating temperature. To elaborate this, from Eq.~(\ref{eq:rhoGW_1}) we obtain the energy density of thermal gravitons in a general cosmological setting as:
\begin{align}
    \rho_{\rm GW}(N_e)=e^{-4 N_e}\int_{N_{e,{\rm ini}}}^{N_e}\mathcal{I}_{\rm GW}(N_e^\prime)\,dN^\prime_e
    \label{eq:rhoGW_2},
\end{align}
where
\begin{align}
\nonumber\mathcal{I}_{\rm GW}(N_e)&\equiv \frac{2\,e^{4N_e}}{\pi^2 } \frac{T^7(N_e)}{M_{\rm pl}^2 H(N_e)} \left(\int_0^\infty \hat k^2 \hat{\eta}\left[T(N_e),\hat k\right]d\hat k \right)\\
&\approx \frac{2\pi^2}{15}\,e^{4N_e}\frac{T^7(N_e)}{M_{\rm pl}^2 H(N_e)}.
\label{eq:IGW}
\end{align}
Here, $N_e$ is the number of e-folds defined by $dN_e=H dt$; throughout we find it convenient to use $N_e$ as a proxy for time.
We assume a vanishing initial graviton energy density at some early time $N_{e,{\rm ini}}$. It is also useful to note that the integrand $\mathcal{I}_{\rm GW}$ represents the rate of change of the comoving GW energy density, i.e. $\mathcal{I_{\rm GW}}\equiv d\left(e^{4N_e} \rho_{\rm GW}\right)/dN_e$. 

From Eq.~\eqref{eq:rhoGW_2}, it is clear that the production of thermal gravitons will depend on the specific details of the WI model under consideration, which determines the corresponding evolution of  $T(N_e)$ and $H(N_e)$.
Nevertheless, we can still extract general conclusions about the GW production within this framework. In fact, $\mathcal{I}_{\rm GW}(N_e)$ is generically an exponentially increasing function through most of the inflationary phase, when $\rho_\phi\gg \rho_r$ and $T$ and $H$ are roughly constant, and it becomes an exponentially decreasing function shortly after the end of inflation at the onset of the radiation-dominated phase, when $\rho_\phi\ll \rho_r$ and $H\propto T^2 \propto e^{-2 N_e}$. The function $\mathcal{I}_{\rm GW}(N_e)$ 
is sharply peaked at some e-fold, $N_e^\rho$, which is the solution of  $d \mathcal{I}_{\rm GW}(N_e) / dN_e = 0$, i.e.,
\bea \left(
4+7\frac{d\ln T}{d N_e}-\frac{d\ln H}{d N_e}\right)_{N_e=N_e^{\rho}}=0.
\label{eq:Nepeak}
\eea
In this sense, $N_e^\rho$ represents the time where most of the contribution to the GW energy density, $\rho_{\rm GW}$, is produced. From that point onward, the $\rho_{\rm GW}$ simply redshifts as $e^{-4N_e}$.  

Thus, at any time subsequent to the epoch of peak graviton production, i.e.  for $N_e > N_e^{\rho}$, 
the thermal graviton yield  expressed in terms of the ratio of energy densities, $\rho_{\rm GW}/\rho_r$, 
can be approximated by using the peak value of $\mathcal{I}_{\rm GW}$ to obtain the following constant value:
\begin{eqnarray}
   \nonumber \frac{\rho_{\rm GW}}{\rho_r}&\simeq&\frac{4}{g_{\star}}\left[\frac{T(N_e^{\rho})}{T(N_e^{\rm RD})}e^{N_e^{\rho}-N_e^{\rm RD}}\right]^4\Delta N_{e}^{\rho} \\
   &\times& \left[\frac{T(N_e^{\rho})}{ H(N_e^{\rho})}\right]\left[\frac{T(N_e^{\rho})}{ \mpl}\right]^2,  \,\,\,\,(N_e>N_e^\rho )
   \label{eq:estrhoratio}
\end{eqnarray}
where $\Delta N_e^{\rho}\gtrsim 1$ denotes the
half-width of $\mathcal{I}_{\rm GW}$ and  $N_e^{\rm{RD}}$ corresponds to the e-fold at which the Universe becomes radiation-dominated, i.e. $N_e^{\rm{RD}}\equiv N_e(\epsilon_H=2)$ where $\epsilon_H\equiv -\dot{H}/H^2$. 
We note that the peak of graviton production takes place in the transition period between the end of inflation ($\epsilon_H =1$) and the onset of radiation domination.
Unless otherwise stated, we assume throughout that 
$g_\star=106.75$, $g_\star(T)=g_{\star,s}(T)$ and $dg_\star(T)/dT=0$. Note that strictly speaking, 
$g_\star$ is set by the specific microphysical realization of WI under consideration. However, its precise value does not qualitatively affect our conclusions, so we set it to its natural SM value.

Since thermal graviton production in WI begins at the onset of the inflationary phase and peaks before the start of the radiation-dominated era, their resulting energy density is expected to be larger than the corresponding value obtained in a scenario with instantaneous reheating into radiation domination  (IRD) at temperature $T_{\rm rh}$, i.e. :
\begin{equation}
   \left[\frac{\rho_{\rm GW}}{\rho_r}\right]_{\rm{IRD},\infty} \simeq \frac{12}{\pi}\sqrt{\frac{10}{g_\star^3}}\left(\frac{T_{\rm rh}}{M_{\rm pl}}\right),\label{eq:dNeff_RD}
\end{equation}
where hereafter the subscript $\infty$ indicates the asymptotic value of a quantity at late times.
To allow a meaningful comparison between these two scenarios, 
 we take the reheating temperature in Eq.~\eqref{eq:dNeff_RD} for the IRD case to correspond to the 
temperature of the thermal bath at the onset of the radiation-dominated phase in the WI scenario, i.e. $T_{\rm rh} = T(N_e^{\rm RD})\equiv T(\epsilon_H = 2)$,  the same as $T(N_e^{\rm RD})$ in  Eq.~\eqref{eq:estrhoratio}.

We introduce the quantity:
\begin{equation}
    R_{\rm GW}\equiv \left[\frac{\rho_{\rm GW}}{\rho_r}\right]_\infty\Big/\left[\frac{\rho_{\rm GW}}{\rho_r}\right]_{\rm IRD,\infty},\label{eq:enhancement_ratio}
\end{equation}
where the numerator refers to the graviton production in WI setting given by Eq.~\eqref{eq:estrhoratio}, and the denominator to the case of instantaneous reheating from Eq.~\eqref{eq:dNeff_RD}. 

In the limit $N_e \rightarrow \infty$, the semi-analytical expression in Eq.~\eqref{eq:estrhoratio} yields the following estimate for $R_{\rm GW}$:
\begin{eqnarray}
    R_{\rm GW}\simeq \frac{\mathcal{I}_{\rm GW} (N_e^{\rho})}{\mathcal{I}_{\rm GW} (N_e^{\rm{RD}})}\Delta N_e^{\rho}. \label{eq:boost_SA}
\end{eqnarray}
Because of the sharp peak feature of $\mathcal{I}_{\rm GW}$, the ratio $\mathcal{I}_{\rm GW} (N_e^{\rho})/\mathcal{I}_{\rm GW} (N_e^{\rm{RD}})$  is  always $\gg1$. As a result, given that $\Delta N_e^{\rho}\gtrsim 1$, we generally find that  $R_{\rm GW}\gg 1$, implying that the thermal graviton production from a WI setting is always enhanced compared to the one in a radiation-dominated phase.

\begin{figure}[ht!]
    \centering
    \includegraphics[width=\linewidth]{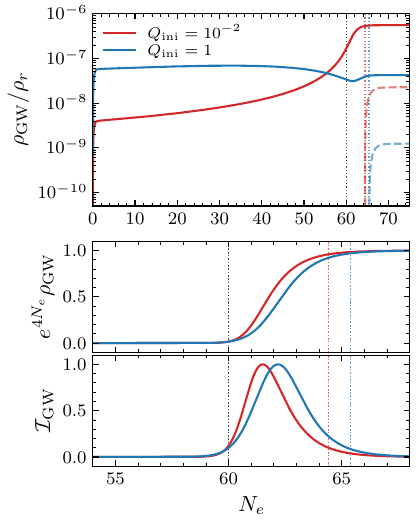}
    \caption{{\it Top Panel:} Evolution of the ratio of the energy density of thermal gravitons, $\rho_{\rm GW}$, to the radiation energy density, $\rho_r$ as a function of the number of e-folds, $N_e$, starting at 60 e-folds before the end of inflation. Solid lines correspond to the WI scenarios with $Q_{\rm ini}=10^{-2}$  (red) and $Q_{\rm ini}=1$ (blue), while dashed lines indicate the corresponding energy density ratios in a radiation-dominated Universe, with the same reheating temperature. The vertical black dotted line marks the end of inflation, whereas the vertical red and blue lines denote the onset of radiation domination respectively for $Q_{\rm ini}=10^{-2}$ and $Q_{\rm ini}=1$. {\it Middle and Bottom Panels:} The comoving GW energy density, $e^{4 N_e} \rho_{\rm GW}$, and its rate of change, $\mathcal{I}_{\rm GW}\equiv d\left( e^{4 N_e} \rho_{\rm GW}\right)/dN_e$, corresponding to the evolution of the energy density shown in the top panel and normalized to have a maximum value of 1. The figure illustrates two key features of thermal graviton production in a WI setting: (i) the substantial enhancement relative to a radiation-dominated Universe, and (ii) its emission is predominantly concentrated in a few e-folds between the end of inflation and the onset of radiation domination. }
    \label{fig:1}
\end{figure}

To validate our results through an explicit case, we consider two representative scenarios characterized by initial dissipation strengths of $Q_{\rm ini} = 10^{-2}$ and $Q_{\rm ini} = 1$, as described at the end of Section~\ref{sec:WIReview}. In Fig.~\ref{fig:1}, we present our numerical results for these two cases. In the top panel, the solid lines represent the ratio of the thermal graviton energy density, $\rho_{\rm GW}$, to the radiation energy density, $\rho_r$, as a function of the number of e-folds, for $Q_{\rm ini} = 10^{-2}$ (red) and $Q_{\rm ini} = 1$ (blue). The dashed lines show the corresponding energy density ratios in a radiation-dominated Universe. The vertical black dotted line indicates the end of inflation, while the vertical red (blue) line marks the onset of radiation domination for $Q_{\rm ini} = 10^{-2}$ ($Q_{\rm ini} = 1$). The bottom panel depicts the evolution of $\mathcal{I}_{\rm GW}$ corresponding to the energy density profiles shown in the top panel, and normalized to have a maximum value of 1. As anticipated, for both WI scenarios under consideration, the graviton production rate reaches its peak shortly before the transition to radiation domination, as marked by the color dotted lines. In addition, the ratios $\rho_{\rm GW}/\rho_r$ rapidly asymptote shortly after entering the radiation-dominated epoch to a final value approximately 30 times greater than the corresponding ones in a radiation-dominated scenario (indicated by the dashed curves), with the same reheating temperature.

\subsection{Spectrum of Gravitational Waves}
Having established the total energy density produced in thermal gravitons, we now turn to characterizing the spectrum of the resulting stochastic GW background. The latter is described by the differential energy density fraction of GWs per logarithmic wavenumber interval, evaluated at e-fold $N_e$: 
\begin{equation}
    \Omega_{\rm GW}\left[k(N_e),N_e\right]\equiv \frac{1}{\rho_c(N_e)}\frac{d\rho_{\rm GW}(N_e)}{d\ln k},
    \label{eq:spectrum}
\end{equation}
where $\rho_c=3 M_{\rm pl}^2 H^2$ is the critical energy density. The time dependence of the physical momentum $k$ is emphasized to highlight the impact of redshifting. The spectrum of the corresponding stochastic GW background from thermal gravitons is obtained using Eqs.~\eqref{eq:spectrum} and \eqref{eq:rhoGW_2}:
\begin{align}
\nonumber    \Omega_{\rm GW}[k(N_e),N_e]&=\frac{e^{-4N_e}}{3 M_{\rm pl}^2H^2(N_e)}\\
    &\times\int_{N_{e,{\rm ini}}}^{N_e} \frac{d\mathcal{I}_{\rm GW}}{d\ln k}[k(N_e^\prime),N_e^\prime]\,dN_e^\prime, \label{eq:Omega_GW0}
\end{align}
where:
\begin{align}
\nonumber    \frac{d\mathcal{I}_{\rm GW}}{d\ln k}[k(N_e),N_e]&=\frac{2\,e^{4N_e}}{\pi^2}\frac{T^7(N_e)}{M_{\rm pl}^2 H(N_e)} \hat k^3 \hat{\eta}\left[T(N_e),\hat k(N_e)\right] \\
    &\approx \frac{15}{\pi^4}\,\mathcal{I}_{\rm GW}(N_e) \frac{\hat k^4}{e^{\hat k}-1}. \label{eq:dIgw_dlnk}
\end{align}
The integrand $d\mathcal{I}_{\rm GW}/d\ln k$, evaluated at a fixed wavenumber $k$, retains the same narrow time profile as $\mathcal{I}_{\rm GW}$. Its maximum occurs at an e-fold we denote $N_e^\Omega$ which lies close to but not exactly at $N_e^\rho$, the peak of $\mathcal{I}_{\rm GW}$. The difference stems from the extra factor 
$\hat k^3 \hat\eta = 
\hat k^4/(e^{\hat k}-1)$ in Eq.~\eqref{eq:dIgw_dlnk}. 
Here, the quantity  $\hat k \equiv k/T$ increases slightly with time during the period of entropy injection prior to radiation domination. The peak of $d\mathcal{I}_{\rm GW}/d\ln k$ thus shifts to earlier or later times depending on whether 
$\hat k(N_e^\rho)$ lies below or above the corresponding $\hat k$ value that maximizes $\hat k^3 \hat\eta$.\footnote{As we explain in the main text, $N_e^\Omega$ depends weakly on the wavenumber $k$; we leave this dependence implicit to streamline the notation.} This additional factor also broadens the peak slightly, giving it a width $\Delta N_e^\Omega\gtrsim \Delta N_e^\rho$ that remains sufficiently narrow for us to approximate the integral in Eq.~\eqref{eq:Omega_GW0} at later times by the product of $d\mathcal{I}_{\rm GW}/d\ln k$ at its peak and $\Delta N_e^\Omega$.

Before evaluating the present-day GW spectrum, it is useful to note that the energy density of gravitons evolves as massless radiation when their wavelength is smaller than the horizon size, whereas modes outside the horizon remain frozen as tensor perturbations until they re-enter during the radiation-dominated epoch after inflation. We define the threshold mode $k_{\rm thr}$ between these two regimes as the mode that exits the horizon at the end of inflation (at e-fold $N_e^{\rm infl}$), i.e. $k_{\rm thr}\equiv H(N_e^{\rm infl})$, which corresponds to a threshold frequency today of $f_{\rm thr}\sim T_0/(2\pi)[H(N_e^{\rm infl})/T(N^{\rm infl}_{e})]$. 

For modes that remain within the horizon at all times, i.e. for frequencies $f>f_{\rm thr}$, Eq.~\eqref{eq:spectrum} can be used to evaluate their spectrum today. However, as we just showed above, the emission of thermal gravitons peaks at very early times and becomes negligible afterward, with the resulting energy density redshifting like that of massless particles. Therefore, for such sub-horizon modes, we compute their present-day spectrum, $\Omega_{\rm GW,0}$, by first evaluating Eq.~\eqref{eq:Omega_GW0} at an arbitrary e-fold, $N^*_{e}$, chosen well within the radiation-dominated era and long after the bulk of the GW production has occured. Subsequently, we redshift the spectrum from $N^*_{e}$ to today according to:
\begin{align}
\nonumber \Omega_{\rm GW,0}[k(N^*_{e})]&=\frac{g_{\star}[T(N^*_{e})]}{g_\star(T_0)}\left(\frac{g_{\star,s}(T_0)}{g_{\star,s}[T(N^*_{e})]}\right)^{\frac{4}{3}}\\
    &\times\Omega_{\gamma,0} \,\Omega_{\rm GW}[k(N^*_{e}), N^*_{e}], \label{eq:Omega_GW0_v2}
\end{align}
where the corresponding GW frequency today is given by:
\begin{equation}
    f=\frac{1}{2\pi}\left(\frac{g_{\star,s}(T_0)}{g_{\star,s}[T(N^*_{e})]}\right)^{\frac{1}{3}}\left(\frac{T_0}{T(N^*_{e})}\right)\,k(N^*_{e}), \label{eq:f_of_k}
\end{equation}
with  $T_0=2.725\,K$, $\Omega_{\gamma,0}=5.38\times 10^{-5}$ is the photon contribution to the total energy density today~\cite{Planck:2018vyg}, $g_\star(T_0)=2$ and $g_{\star, s}(T_0)=3.931$, respectively.
Combining Eqs.~\eqref{eq:Omega_GW0} through \eqref{eq:f_of_k}, and exploiting the sharply peaked nature of $d\mathcal{I}_{\rm GW}/d\ln k$ around $N_e^{\Omega}$ (we approximate the integral in Eq.~\eqref{eq:Omega_GW0} as the product of $d\mathcal{I}_{\rm GW}/d\ln k$ at its peak times $\Delta N_e^\Omega$), we obtain the following expression for the present‑day spectrum of modes that remain subhorizon throughout:
\begin{widetext}
\begin{align}
    \Omega_{\rm GW,0}(f)&\simeq \frac{960 }{g_{\star}(T_0)}\Omega_\gamma\,\Delta N_{e}^{\Omega}\,\left[\frac{T^3(N_e^{\Omega})}{ H(N_e^{\Omega}) \mpl^2}\right] \left(\frac{f}{T_0}\right)^4\left\{\exp\left[2\pi\left(\frac{g_{\star,s}[T(N^*_{e})]}{g_{\star,s}(T_0)}\right)^{\frac{1}{3}}\left( \frac{T(N_e^*)e^{N_e^*}}{T(N_e^\Omega)e^{N_e^\Omega}}\right)\frac{f}{T_0}\right]-1\right\}^{-1},\,\,f> f_{\rm thr}.\label{eq:Omega_GW0_of_f_subhor}
\end{align}
\end{widetext}
As expected, here we see explicitly that once the GW spectrum is re‑expressed in terms of the present‑day frequency $f$, there is effectively no dependence on the choice of e-fold reference $N_e^*$. In fact, the product $T(N_e^*)e^{N_e^*}$ is constant for any choice of $N_e^*$ due to entropy conservation from the onset of radiation domination onward. The only residual and very mild dependence is in the factor $(g_{\star,s}(T_0)/g_{\star,s}[T(N_e^*)])^{1/3}$ appearing in the exponential in the last term of Eq.~\eqref{eq:Omega_GW0_of_f_subhor}.
In our analysis, for simplicity,  this factor is 1 since we have taken $g_\star=g_{\star,s} = 106.75$ for all times prior to the reference e-fold $N_e^\star$, i.e. for all $N_e\leq N_e^*$.

We now turn to modes that exit the horizon during inflation and re-enter in the radiation-dominated era, corresponding to frequencies $f\leq f_{\rm thr}$. For these modes the evolution up to horizon exit is still governed by Eq.~\eqref{eq:Omega_GW0}. Integrating this equation up to the e-fold $N_e^{\rm ex}$ at which $k=H$ yields the frozen power spectrum of thermal tensor fluctuations:
\begin{equation}
\mathcal{P}^{\rm th}_T(k_c)=12\,\Omega_{\rm GW}[k(N_e^{\rm ex}),N_e^{\rm ex}], \label{eq:Pth}
\end{equation}
where $k_c\equiv e^{N_e}\,k$ is the comoving wavenumber and the superscript ``th'' denotes the thermal origin of the spectrum. We treat $\mathcal{P}^{\rm th}_T(k_c)$ as an additional source of primordial GWs, alongside the standard vacuum power tensor spectrum sourced by inflaton fluctuations:
\begin{equation}
\mathcal{P}^{\rm vac}_T(k_c)=2\left[H^2(N_e^{\rm ex})\right]/(\pi^2 M_{\rm pl}^2). \label{eq:standardPower} 
\end{equation}
 In this sense, we can evaluate the thermal contribution to the present-day thermal GW spectrum for these modes  as is commonly done for primordial tensor fluctuations in the literature, such that~\cite{Kolb:1990vq}:
\begin{align}
\nonumber    \Omega_{\rm GW,0}([k(N^{\rm re}_{e})])&=\frac{1}{24}\frac{g_{\star}[T(N^{\rm re}_{e})]}{g_\star(T_0)}\\
 &\times\left(\frac{g_{\star,s}(T_0)}{g_{\star,s}[T(N^{\rm re}_{e})]}\right)^{\frac{4}{3}} \Omega_{\gamma,0}\,\mathcal{P}^{\rm th}_T(k_c), \label{eq:Omega_GW0_v3}
\end{align}
where $N_e^{\rm re}$ denotes the e-fold at which the mode with comoving wavenumber $k_c$ re-enters the horizon during radiation domination, and corresponding GW frequency today given by:
\begin{equation}
    f=\frac{1}{2\pi}\left(\frac{g_{\star,s}(T_0)}{g_{\star,s}[T(N^{\rm re}_{e})]}\right)^{\frac{1}{3}}\left(\frac{T_0}{T(N^{\rm re}_{e})}\right)\,k(N_e^{\rm re}). \label{eq:f_of_k_v2}
\end{equation}
To evaluate $\mathcal{P}_T^{\rm th}(k_c)$ using Eq.~\eqref{eq:Omega_GW0}, we first recall that $d\mathcal{I}_{\rm GW}/d\ln k$ grows exponentially throughout inflation, and therefore reaches its largest value at horizon exit. The integral in Eq.~\eqref{eq:Omega_GW0} is thus dominated by its upper limit, allowing us to approximate it by simply evaluating the integrand at 
$N_e^{\rm ex}$. Combining Eqs.~\eqref{eq:Pth}
through \eqref{eq:f_of_k_v2} with this observation yields a compact expression for the present-day GW spectrum for modes that exit the horizon during inflation and re-enter in the radiation era:
\begin{widetext}
\begin{equation}
    \Omega_{\rm GW, 0}(f)\simeq \frac{\Omega_\gamma}{3\pi^2}\frac{g_{\star}[T(N^{\rm re}_{e})]}{g_\star(T_0)}\,\left(\frac{g_{\star,s}(T_0)}{g_{\star,s}[T(N^{\rm re}_{e})]}\right)^{\frac{4}{3}} \left[\frac{T(N^{\rm ex}_{e})}{M_{\rm pl}}\right]^4 ,\,\, f\leq f_{\rm thr}.\label{eq:Omega_GW0_of_f_superhor}
\end{equation}
\end{widetext}
Here we used the fact that a WI setting must satisfy the condition $H<T$ throughout.
Hence at horizon crossing, where $k=H$, $\hat{k}= H/T\ll 1$ so that we may approximate $[\hat k^3 \hat \eta](N_e^{\rm ex})\approx \hat{k}^3(N_e^{\rm ex})=[H(N_e^{\rm ex})/T(N_e^{\rm ex})]^3$.

The two key expressions derived above -- namely Eqs.~\eqref{eq:Omega_GW0_of_f_subhor} and~\eqref{eq:Omega_GW0_of_f_superhor} -- completely characterize the present-day thermal graviton spectrum in WI scenarios. Together, these equations reveal a distinctive spectral shape with two primary features: a sharp peak at high frequencies, of the order of the CMB temperature today, and an approximately flat plateau extending to lower frequencies. 

The high-frequency peak arises from the sub-horizon modes. Such peak is expected to be higher than the corresponding thermal graviton spectrum from a radiation-dominated era with the same reheating temperature, reflecting the same enhancement factor introduced in the context of the total GW energy density, i.e. $R_{\rm GW}$ in Eq.~\eqref{eq:boost_SA}. There is also a mild peak-frequency shift relative to the GW spectrum from a radiation-dominated era due to the factor  $[T(N_e^*)e^{N_e^*}]/[T(N_e^\Omega)e^{N_e^\Omega}]$ in Eq.~\eqref{eq:Omega_GW0_of_f_subhor}. In fact, while this term equals unity in a radiation-dominated Universe, in a WI setting the non-zero entropy injection prior to radiation domination makes it slightly larger than one, shifting the peak of the GW spectrum to lower frequencies. However, as we will show below, this effect is negligible in practice. 

The low-frequency plateau corresponds instead to the horizon-crossing modes that exited the horizon during WI, became frozen as tensor perturbations, and subsequently re-entered during radiation domination. This flat spectral region directly reflects the approximately constant temperature evolution characteristic of a WI setting. The threshold frequency $f_{\rm thr}$ marks the boundary between these peaked and flat spectral regions, corresponding to the mode that is equal to the horizon size at the end of inflation.

\begin{figure}
    \centering
    \includegraphics[width=\linewidth]{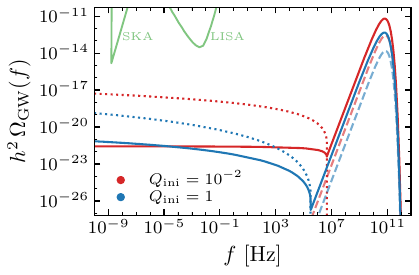}
    \caption{Spectrum of stochastic GWs from thermal gravitons in WI scenarios characterized by initial dissipation strengths of $Q_{\rm ini} = 10^{-2}$ (red) and $Q_{\rm ini}= 1$ (blue). Solid lines depict the thermal graviton contribution from WI, while dashed lines show the corresponding thermal graviton spectra from a radiation-dominated epoch with matching reheating temperatures. Dotted lines display the spectrum of standard vacuum tensor modes sourced by inflaton fluctuations; these vacuum tensor modes are seen to dominate over the thermal graviton contribution at the low frequency plateau. The spectra are shown as $h^2\Omega_{\rm GW}$ with $h = 0.68$. Finally, for reference we indicate in green the projected sensitivity limits from SKA~\cite{Janssen:2015ft} and LISA~\cite{LISA:2017pwj}. The figure illustrates the key spectral features of the thermal GW spectrum from WI: (i) the characteristic shape with a sharp high-frequency peak and a flat low-frequency plateau; (ii) the enhanced peak amplitudes in WI settings relative to radiation-dominated cases; and (iii) the challenging detection prospects, as the thermal graviton signals lie well below the sensitivity thresholds of planned observational surveys. }
    \label{fig:OmegaGW_1}
\end{figure}

Fig.~\ref{fig:OmegaGW_1} shows our calculated thermal graviton spectra. Specifically, the solid red (blue) line represents the spectrum of thermal gravitons in a WI setting with initial dissipation strength of $Q_{\rm ini}=10^{-2}$ ($Q_{\rm ini}=1$). The dashed lines depict instead the spectrum of thermal gravitons from a
radiation-dominated era with the same reheating temperature
as the WI scenarios. Overall, the results displayed in the figure confirm the analytical expectations discussed above. Namely, we observe: (i) the characteristic spectral shape with a sharp high-frequency peak and an approximately flat low-frequency plateau; (ii) an enhanced peak amplitude in the WI scenarios relative to radiation-dominated cases, along with a non-zero but negligible frequency shift of the peak toward lower frequencies. 

In addition to the WI thermal  gravitons discussed above, we must also consider the standard vacuum tensor modes sourced by inflation fluctuations.\footnote{The present-day spectrum of vacuum tensor modes can be obtained following the same procedure as the thermal spectrum, starting from Eq.~\eqref{eq:Omega_GW0_v3} and substituting $\mathcal{P}_T^{\rm vac}$ instead of $\mathcal{P}_T^{\rm th}$.} 
We display the latter as dotted lines in Fig.~\ref{fig:OmegaGW_1}. Notably, these vacuum tensor contributions dominate over the thermal contributions for horizon-crossing modes (the flat plateau part of the spectrum) in the representative WI scenarios analyzed here. It is important to note that the vacuum tensor fluctuations differ between the two representative WI examples shown, as the models require different potential heights to satisfy observational constraints on the scalar power spectrum from CMB measurements (for details see Ref.~\cite{Freese:2024ogj}). In this sense, the relative importance of the vacuum to thermal contributions is evidently model-dependent, and we discuss it in further detail in the next section, specifically in the context of current observational constraints and future detection prospects for the thermal graviton spectrum, and how these relate to the evolution of underlying temperature and Hubble expansion rate.

\section{Observational Implications and Constraints}
\label{sec:constraints}

Having established the total energy density and spectral characteristics of thermal gravitons in WI, we now turn to their observational implications. These signatures can be probed through two complementary channels: direct constraints from the GW spectrum itself, and indirect constraints from the contribution to the effective number of relativistic species $N_{\rm eff}$. While the former provides frequency-dependent information about the underlying physics, the latter offers a model-independent probe of the total energy density in thermal gravitons. Together, these observational avenues allow us to map the viable parameter space for WI scenarios and assess the prospects for a future detection of its characteristic thermal GW background.

\subsection{Constraints and Detection Prospects for the thermal GW spectrum}

\begin{figure}[ht!]
    \centering
    \includegraphics[width=\linewidth]{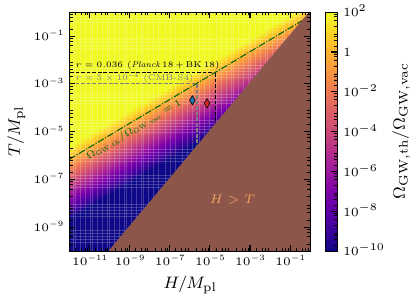}
    \caption{Ratio of the low-frequency thermal and vacuum GW spectra, $\Omega_{\rm GW,0}^{\rm th}/\Omega_{\rm GW,0}^{\rm vac}$ as a function of the Hubble parameter $H$ and temperature $T$ during inflation. The color scale to the right indicates the magnitude of $\Omega_{\rm GW,0}^{\rm th}/\Omega_{\rm GW,0}^{\rm vac}$. The dot-dashed green line in the figure  marks the boundary at which the thermal contributions equal those from vacuum tensor perturbations; in the yellow region above the line, the thermal contribution dominates. The brown shaded region corresponds to $H > T$, where WI is not viable. Black and gray dashed lines represent current and future constraints on the tensor-to-scalar ratio $r$ at $2\sigma$, respectively from the combination of $\textit{Planck}$ and BICEP/\textit{Keck} measurements~\cite{BICEP:2021xfz}, and CMB-S4 forecasts~\cite{CMB-S4:2020lpa}. Note that these constraints apply specifically to CMB modes, which exit the horizon approximately 60 e-folds before the end of inflation. The region of particular interest lies between the green contour and the constraint lines, where thermal contributions dominate over vacuum ones while remaining consistent with current observational bounds and within reach of future experiments such as CMB-S4. For reference, we indicate as red ($Q_{\rm ini} = 10^{-2}$) and blue ($Q_{\rm ini}= 1$)  diamonds the corresponding values of $\Omega_{\rm GW,0}^{\rm th}/\Omega_{\rm GW,0}^{\rm vac}$ for the two representative WI examples analyzed in this work. As a whole, the figure highlights the fundamental challenge of observationally distinguishing WI signatures through thermal GWs in realistic scenarios, requiring ``fine-tuned'' combinations of $H,\,T$ to achieve both thermal dominance and observational accessibility.}
    \label{fig:OmegaGW_2}
\end{figure}

We begin by examining the observational prospects for detecting the thermal graviton spectrum and the constraints it places on WI parameters.

The enhanced high-frequency peak presents an interesting target for detection efforts, though with challenging prospects. In recent years, searches for high-frequency GWs above the kHz range have attracted vast interest (see~\cite{Aggarwal:2025noe} for a review of recent developments).   Search strategies typically rely on the inverse Gertsenshtein effect~\cite{gertsenshtein1962wave}, where an effective conversion of GWs into electromagnetic fields occurs as they propagate through an external magnetic field. However, the probability of graviton conversion is suppressed by $M^2_{\rm pl}$. Nevertheless exclusion limits have been set based on the non-detection of conversion in the magnetic fields in astrophysical environments \cite{Domcke:2020yzq, Liu:2023mll, Ito:2023fcr, Ito:2023nkq, He:2023xoh, Dandoy:2024oqg, Lella:2024dus, McDonald:2024nxj, Dunsky:2025pvd, Matsuo:2025blj}, as well as in existing and proposed laboratory facilities \cite{Ejlli:2019bqj, Domcke:2022rgu, Gao:2023gph, Domcke:2023bat, Valero:2024ncz, Domcke:2024mfu, Domcke:2024eti, Capdevilla:2024cby, Pappas:2025zld}.  Ref.~\cite{Ringwald:2020ist} has shown that, even under optimistic projections using the Gaussian beam technique in future graviton/photon conversion experiments, a reheating temperature near the Planck scale is required in the case of instantaneous reheating into a radiation-dominated universe.  In the context of thermal graviton detection,  the thermal GW signal we derived here within a WI setting only offers a modest improvement over this challenging baseline, with an enhancement factor $R_{\rm GW}\sim 30$ relative to instantaneous reheating scenarios to a radiation-dominated era with the same initial temperature. Consequently, the detection prospects of this high-frequency signal ultimately depend on the viability to achieve a sufficiently high reheating temperature within a realistic WI framework. Given the well-defined peak frequency of the thermal GW spectrum, a complementary strategy is to employ dedicated resonant searches targeting the signal around $10^{11}\,{\rm Hz}$. For instance, Ref.~\cite{Herman:2022fau} explores a resonant cavity design capable of reaching a characteristic strain sensitivity of $\mathcal{O}(10^{-36})$ at $10^{8}\,{\rm Hz}$. With further optimization toward the THz band, such resonant setups could provide a promising avenue to probe WI thermal gravitons. Below we describe the bounds set by $\Delta N_{\rm eff}$ measurements on the graviton energy density, which are far more constraining than any existing bounds from graviton/photon conversion.

Given these practical limitations in observing the high-frequency regime, it is also important to consider the low-frequency plateau of the GW spectrum, corresponding to horizon-crossing modes. In this region, the quantity of interest is the relative importance of thermal versus vacuum tensor contributions, which depends critically on the temperature and Hubble parameter evolution during WI, namely:
\begin{equation}
    \frac{\Omega_{\rm GW,0}^{\rm th}}{\Omega_{\rm GW,0}^{\rm vac}}\equiv \frac{\mathcal{P}^{\rm th}_T}{\mathcal{P}_T^{\rm vac}}\simeq 4\,\left[\frac{T(N_e^{\rm ex})}{H(N_e^{\rm ex})}\right]^2\,\left[\frac{T(N_e^{\rm ex})}{M_{\rm pl}}\right]^2, \label{eq:OGWth_over_OGWvac}
\end{equation}
where we refer to the low-frequency thermal and vacuum gravitational spectra today respectively as $\Omega_{\rm GW,0}^{\rm th}$ and $\Omega_{\rm GW,0}^{\rm vac}$. Note that the ratio of spectra on the left hand side of Eq.~\eqref{eq:OGWth_over_OGWvac} depends implicitly on the comoving wavenumber $k_c$, since each mode exits the horizon at a different e-fold, $N_e^{\rm ex}$. While WI is characterized by $T>H$ throughout, the factor $(T/M_{\rm pl})^2$ introduces a strong suppression that is  typically greater than the boost $(T/H)^2$, keeping the thermal contributions subdominant. This is clearly illustrated in Fig.~\ref{fig:OmegaGW_1}, where we observe that for the two representative WI examples analyzed here, the thermal contributions remain consistently below the vacuum tensor spectrum. This relationship is a typical feature in single-field slow-roll WI models. For the case of a monomial inflaton potential $V(\phi)=\lambda_n M_{\rm pl}^4\,\left(\phi/M_{\rm pl}\right)^n$ and a constant dissipation strength $Q$, we find that~(see Appendix~\ref{A:1} for the full derivation):
\begin{align}
    \frac{\Omega_{\rm GW,0}^{\rm th}}{\Omega_{\rm GW,0}^{\rm vac}}\Big|_{\rm mWI}&\approx 2.0\times 10^{-6}\, \left( \frac{106.75}{g_\star[T(N_e^{\rm ex})]}\right)^4 \nonumber \\
    &\times\,\frac{1}{n\left(n+4\,\Delta N_e^{\rm ex}\right)}\frac{Q}{1+Q},
\end{align}
where $\Delta N_e^{\rm ex}\equiv(N^{\rm infl}_e -N_e^{\rm ex})>0$, with $N_e^{\rm infl}=60$.\footnote{Here $N_e^{\rm infl}=60$ denotes the end of inflation, and we set initial conditions at $N_e^{\rm ex}=0$, corresponding approximately to the horizon exit of CMB modes. Note that this choice of e-fold counting is arbitrary and does not affect physical results.} The  factor in the second line is evidently always less than unity for
$n \gtrsim \mathcal{O}(1)$, thus ensuring that the thermal contributions to the primordial spectrum remain highly suppressed in minimal WI models.

Despite this general suppression in the simplest WI scenarios, we adopt a broader approach by exploring the full parameter space to identify combinations of temperature and Hubble parameter values for which thermal contributions could dominate over the standard vacuum tensor spectrum. Fig.~\ref{fig:OmegaGW_2} presents this two-dimensional parameter space analysis, showing the ratio of thermal to vacuum tensor contributions as a function of $H$ and $T$ during inflation. The color-coding of the value of the ratio is indicated on the right.  The green dot-dashed line delineates the boundary between the case where the thermal graviton contribution dominates (above the line) vs.
where the vacuum contribution dominates (below the line).  The two WI scenarios analyzed in this work are indicated by diamonds, in red and blue for $Q_{\rm ini} = 10^{-2}$ and $Q_{\rm ini}= 1$, respectively.

In Fig.~\ref{fig:OmegaGW_2}, we also illustrate the observational constraints from CMB data on the tensor-to-scalar ratio $r$.
This ratio is calculated by normalizing the obtained tensor spectra $\mathcal{P}_T^{\rm th}$ and $\mathcal{P}_T^{\rm vac}$, for a given pair of $H(N_e^{\rm ex})$ and $T(N_e^{\rm ex})$ values, to the observed primordial scalar power spectrum amplitude $A_s\approx 2.1\times 10^{-9}$ from CMB measurements~\cite{Planck:2018vyg}.
Black and gray dashed lines represent current and future constraints on $r$ at $2\sigma$, respectively, from the combination of $\textit{Planck}$ and BICEP/\textit{Keck} measurements~\cite{BICEP:2021xfz}, and CMB-S4 forecasts~\cite{CMB-S4:2020lpa}. This figure thus demonstrates how existing primordial GW limits already restrict the conditions under which thermal signatures could be observable relative to both the vacuum contribution and current/near-future experimental sensitivities. 

Specifically, for $H\lesssim 5\times10^{-6}\,M_{\rm pl}$ and $T\sim 10^{-3}\,M_{\rm pl}$, we can see that thermal contributions can surpass the vacuum tensor spectrum while generating a tensor-to-scalar ratio that remains below current observational bounds, yet falls within the projected sensitivity of future experiments such as CMB-S4. This represents an optimal window where thermal signatures could be both dominant and detectable. However, for lower temperatures $T \lesssim 10^{-3} M_{\rm pl}$, while a substantial region of parameter space still exists where thermal contributions exceed vacuum ones, the overall amplitude of the resulting GW spectrum becomes too weak to be observationally accessible with foreseeable experimental capabilities. In summary, this analysis highlights the fundamental challenge of observationally distinguishing WI signatures through thermal GWs, with a detection that would require not only the right balance of inflationary parameters to achieve thermal dominance, but also sufficiently high energy scales to produce an observable signal.

\subsection{Dark Radiation Constraints}

As discussed in the previous section, direct detection prospects for thermal gravitons from WI remain challenging across both low and high frequency ranges. That said, the thermal gravitons emitted during WI can also be indirectly constrained by cosmological observables. Specifically, the thermal graviton background  contributes to dark radiation, manifesting as an increase in the effective number of relativistic species $N_{\rm eff}$ that can be precisely measured through its impact on both BBN and more importantly the CMB. The contribution of thermal gravitons to $\Delta N_{\rm eff}$ it simply given by Eq.~\eqref{eq:estrhoratio}, rescaled by by the neutrino-to-photon normalization factor $a_\nu\equiv 8/7\,\left(11/4\right)^{4/3}\approx 4.40$, i.e.:
\begin{align}
    \Delta N_{\rm eff}&=a_\nu\, \left[\frac{\rho_{\rm GW}}{\rho_r}\right]_\infty \nonumber \\
    &\simeq 
    \frac{4\, a_\nu}{g_{\star}}\left[\frac{T(N_e^{\rho})}{T(N_e^{\rm RD})}e^{N_e^{\rho}-N_e^{\rm RD}}\right]^4 \nonumber \\
   &\times\Delta N_{e}^{\rho}\left[\frac{T(N_e^{\rho})}{ H(N_e^{\rho})}\right]\left[\frac{T(N_e^{\rho})}{ \mpl}\right]^2. 
   \label{eq:estY2}
\end{align}

While the precise value of $\Delta N_{\rm eff}$ depends on the detailed evolution between the peak in GW energy density production, at $N_e^\rho$, and the onset of radiation domination, at $N_e^{\rm RD}$, we can estimate a conservative upper bound by setting $[T(N_e^{\rho})/T(N_e^{\rm RD})\cdot e^{N_e^{\rho}-N_e^{\rm RD}}]=1$. This approximation assumes a smooth transition to the radiation-dominated phase after the end of inflation, where both the bath temperature and the Hubble parameter are monotonically decreasing functions of e-folds. In these simple scenarios,  the temperature evolves more gradually than in radiation domination due to the ongoing entropy injection from the inflaton decay, which slows the cooling relative to standard adiabatic expansion.

\begin{figure}
    \centering
    \includegraphics[width=\linewidth]{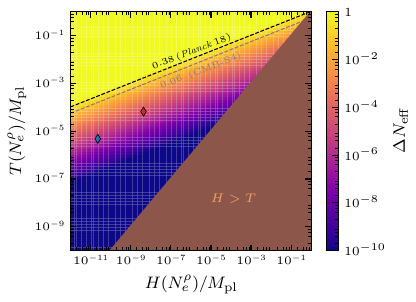}
    \caption{The contribution to the effective number of relativistic species, $\Delta N_{\rm eff}$, from thermal gravitons as a function of the Hubble parameter $H$ and temperature $T$ at the peak of the GW production, $N_e^\rho$. The color scale to the right indicates the magnitude of $\Delta N_{\rm eff}$,  with the black and gray dashed lines marking the corresponding current and forecasted constraints, respectively from \textit{Planck}~\cite{Planck:2018vyg} and CMB-S4~\cite{CMB-S4:2016ple}, at the $2\sigma$ level. The brown shaded region corresponds to $H > T$, where WI is not viable.  The figure demonstrates that the $\Delta N_{\rm eff}$ contributions from thermal gravitons remain negligible across the vast majority of physically motivated parameter space, further underlined by the small values obtained for the two representative WI examples analyzed in this work, shown as red ($Q_{\rm ini} = 10^{-2}$) and blue ($Q_{\rm ini}= 1$) diamonds. }
    \label{fig:2}
\end{figure}

Under this conservative assumption, we can map the parameter space of temperature and Hubble parameter values that yield observationally significant $\Delta N_{\rm eff}$ contributions, which we illustrate in Fig.~\ref{fig:2}. There, we display the two-dimensional parameter space analysis, showing the $\Delta N_{\rm eff}$ contributions from thermal gravitons as a function of $H$ and $T$ at the peak of the graviton production, $N_e^\rho$, alongside current and forecasted observational constraints on $\Delta N_{\rm eff}$ from CMB experiments. As a whole, the figure demonstrates that the thermal GW contributions to dark radiation are negligible across a large portion of the parameter space. This is exemplified by the blue and red diamonds representing the two WI examples analyzed in this work, which yield $\Delta N_{\rm eff}$ values well below $10^{-6}$, orders of magnitude below even future observational sensitivity. The only scenarios that are constrained by current $\Delta N_{\rm eff}$ measurements require exceptionally high temperatures, $T\gtrsim10^{-3} M_{\rm pl}$, combined with extreme temperature-to-Hubble ratios of $T/H\gtrsim 10^{6}$. Such parameter combinations are unrealistic within standard WI frameworks, as they would necessitate a contrived post-inflationary evolution where temperatures increase rather than decrease -- a scenario that would likely require additional exotic mechanisms beyond the minimal WI setup. In fact, for the simple case of monomial inflation potential and a constant dissipation strength $Q$, we can derive strong constraints on the maximum $\Delta N_{\rm eff}$ contribution, namely~(see Appendix~\ref{A:2} for the full derivation):
\begin{align}
\Delta N_{\rm eff}&< 2.1\times 10^{-7} \left( \frac{106.75}{g_\star[T(N_e^{\rm ex})]}\right)^4\,\frac{Q^{\frac{3}{4}}}{1+Q} \nonumber \\
&\times \frac{1}{n^{\frac{3}{2}}\left(1+4\,N_e^{\rm cmb}/n\right)^{\frac{n+2}{8}}},
\end{align}
where the last two multiplicative factors are always less than unity for $n \gtrsim \mathcal{O}(1)$, confirming that $\Delta N_{\rm eff}$ contributions remain strongly suppressed in realistic WI scenarios.

Collectively, these results establish that while thermal gravitons represent a distinctive signature of WI, their contribution to dark radiation provides neither a viable detection channel nor a meaningful constraint on realistic WI scenarios, highlighting once again the observational challenges inherent in distinguishing WI from standard cold inflation through GW signatures.  

\medskip
\section{CONCLUSIONS}
\label{sec:conclusions}
In this work, we have explored the emission and evolution of high-energy thermal gravitons generated by collisions in the thermal bath during WI.

We showed that most of the thermal graviton production happens before the transition from vacuum to radiation domination after the end of inflation is complete. We also demonstrated that the energy density of emitted GWs is roughly one to two orders of magnitude enhanced compared to that of the GWs produced during radiation domination with the same reheating temperature. By computing the spectrum of the resulting stochastic GW background, we found that it exhibits a unique shape: a sharp peak at high frequencies ($\sim 100\,{\rm GHz}$ which is of the order of the CMB temperature today) as well as a nearly flat plateau stretching to low frequencies. The peak in the spectrum is attributed to sub-horizon modes that follow
the temperature of the bath, while the flat part arises from the modes that become super-horizon during WI and re-enter during radiation domination.

Additionally, we examined the detectability of the thermal stochastic GW background generated by WI and concluded that the high-frequency peak, despite being enhanced compared to the radiation-dominated scenario, remains difficult to observe unless the temperature is sufficiently close to the Planck scale. We also showed that although the low-frequency plateau of the thermal spectrum is generally expected to remain subdominant to the vacuum tensor spectrum, it is not impossible for WI models to exist in which the thermal spectrum can exceed the vacuum contribution. To this end, we investigated the conditions on the evolution of the temperature and the Hubble expansion rate during WI that would allow the low-frequency plateau to surpass the vacuum tensor spectrum while maintaining the tensor-to-scalar ratio within current observational bounds. Finally, we computed the contribution of thermal gravitons produced by WI to dark radiation and demonstrated that WI models are typically consistent with current observational bounds on $\Delta N_{{\rm eff}}$ from the CMB, and  constraints would only arise for models with exceptionally high temperatures and extreme temperature-to-Hubble rate ratios.

Despite the observational challenges, the irreducible thermal gravitons produced during WI provide a distinctive and appealing feature that motivates further investigation into more possible realizations of this class of models and their potential observational signatures. To facilitate such investigations, the parameter space maps presented in Figs.~\ref{fig:OmegaGW_2} and~\ref{fig:2} offer a systematic framework that allows theorists to directly evaluate  the observational viability of specific WI models, based on their predicted Hubble and temperature evolution.

\acknowledgments

We thank Kun-Feng Lyu for useful conversations. K.F., G.M. and B.S.E. are grateful for support from
the Jeff \& Gail Kodosky Endowed Chair in Physics at the University of Texas. K.F. G.M. and B.S.E. acknowledge support
by the U.S. Department of Energy, Office of Science, Office of High Energy Physics program under Award Number DE-SC-0022021. K.F. and G.M. also acknowledge
support from the Swedish Research Council (Contract
No. 638-2013-8993).  T.X. is supported by the U.S. National Science Foundation under Award No. PHY-2412671.

\renewcommand\thesubsection{\arabic{subsection}}

\appendix
\section{Constraints on the thermal GW spectrum in a minimal WI setting}
In this appendix, we derive analytical estimates for the thermal GW signatures in the simplest WI scenarios. Specifically, we consider a minimal setup where the inflationary phase is described by single-field slow-roll dynamics with a monomial potential $V(\phi) = \lambda_n M_{\rm pl}^4\left(\phi/M_{\rm pl}\right)^n$ with constant dissipation strength $Q$, and where the radiation-dominated phase is reached smoothly after the end of inflation. Under these assumptions, we obtain analytical expressions for both the contribution to dark radiation, $\Delta N_{\rm eff}$, and the ratio of thermal to vacuum contributions in the GW spectrum, $\Omega_{\rm GW, 0}^{\rm th}/\Omega_{\rm GW. 0}^{\rm vac}$. While simplified, this framework captures the essential physics of WI, and provides a concrete baseline for understanding the observational prospects of thermal GW signatures in realistic WI constructions.
\label{sec:appA}
\subsection{An upper bound on the contribution of thermal GWs to the primordial tensor spectrum}\label{A:1}

To establish an upper bound on the thermal GW contribution in a slow-roll monomial WI, we begin by specifying the corresponding inflationary dynamics, namely:
\begin{align}
    H&\approx \frac{V}{3M_{\rm pl}^2}=\sqrt{\frac{\lambda_n}{3}}M_{\rm pl}\tilde{\phi}^{\frac{n}{2}}, \label{eq:A1}\\
    \dot{\phi}^2&\approx -\frac{V_{,\phi}}{3H(1+Q)}=-\sqrt{\frac{\lambda_n}{3}}\frac{M_{\rm pl}^2}{1+Q}\tilde\phi^{\frac{n-2}{2}}, \label{eq:A2} \\
    T&\approx \frac{30}{\pi^2 g_\star}\left(\frac{3Q}{4}\dot\phi^2\right)^{\frac{1}{4}} \nonumber \\
    &= \frac{30}{\pi^2 g_\star} \left[\frac{\lambda_n Q}{4(1+Q)^2}\right]^{\frac{1}{4}}M_{\rm pl}\tilde\phi^{\frac{n-2}{4}}, \label{eq:A3}
\end{align}
where $\tilde\phi\equiv \phi/M_{\rm pl}$. The field value at the end of inflation $\tilde\phi_{\rm end}$ is determined by setting the slow-roll parameter $\epsilon\approx M_{\rm pl}^2\left(V_{,\phi}/V\right)^2/[2(1+Q)]$ to unity, yielding:
\begin{equation}
    \tilde\phi_{\rm end}\approx \frac{n}{\sqrt{2(1+Q)}}.\label{eq:A4}
\end{equation}
From this relation, we can determine the field value at any e-fold $N_e^{\rm ex}$ prior to the end of inflation by evaluating the following integral:
\begin{equation}
    \Delta N_e^{\rm ex}=\frac{1+Q}{M_{\rm pl}}\int_{\tilde\phi(N_e^{\rm ex})}^{\tilde\phi_{\rm end}}\frac{V_{,\phi}}{V}d\tilde\phi, \label{eq:A5}
\end{equation}
where $\Delta N_e^{\rm ex}\equiv N_e^{\rm infl}-N_e^{\rm ex}$. Here, $N_e^{\rm infl}=60$ marks the end of inflation, and $N_e^{\rm ex}=0$ corresponds approximately to when the CMB modes relevant to current observations first exited the horizon. Note that this choice of e-fold counting is arbitrary and serves only as a convenient reference point, having no effect on physical results.

Combining Eqs~.\eqref{eq:A1},~\eqref{eq:A3},~\eqref{eq:A4} and~\eqref{eq:A5} allows us to obtain the values of the thermal bath temperature $T$ and Hubble parameter $H$ at $N_e^{\rm ex}$:
\begin{align}
    H(N_e^{\rm ex})&\approx \sqrt{\frac{\lambda_n}{3}}\left[\frac{n^2}{4(1+Q)}\,\left(1+\frac{4\Delta N_e^{\rm ex}}{n}\right)\right]^{\frac{n}{4}}\,M_{\rm pl}, \label{eq:A6}\\
    T(N_e^{\rm ex})&\approx \frac{30}{\pi^2g_\star}\left[\frac{\lambda_n Q}{2^n(1+Q)^2}\right]^{\frac{1}{4}}\nonumber\\
    &\times \left[\frac{n^2}{1+Q}\,\left(1+\frac{4\Delta N_e^{\rm ex}}{n}\right)\right]^{\frac{n-2}{8}}\,M_{\rm pl}. \label{eq:A7}
\end{align}
Finally, we can substitute the above expressions in Eq~\eqref{eq:OGWth_over_OGWvac} to determine the ratio of thermal to vacuum contributions in the primordial GW spectrum:
\begin{align}
    \frac{\Omega_{\rm GW,0}^{\rm th}}{\Omega_{\rm GW,0}^{\rm vac}}&\approx 12\left(\frac{30}{\pi^2 g_\star}\right)^4\frac{1}{n\left(1+4\Delta N_e^{\rm ex}\right)}\frac{Q}{1+Q} \\
    &\approx 2.0\times 10^{-6}\, \left( \frac{106.75}{g_\star[T(N_e^{\rm ex})]}\right)^4 \nonumber \\
    &\times\,\frac{1}{n\left(n+4\,\Delta N_e^{\rm ex}\right)}\frac{Q}{1+Q}.
\end{align}

This result demonstrates that the thermal GW contribution is always subdominant to the vacuum contribution in slow-roll monomial WI scenarios. In fact, the ratio $\Omega_{\rm GW,0}^{\rm th}/\Omega_{\rm GW,0}^{\rm vac}$ is evidently $\ll 1$ for all values of $Q$ and $n \gtrsim \mathcal{O}(1)$, with the suppression becoming moderately more pronounced for steeper potentials (larger $n$) and in a strong dissipative regime ($Q\gg 1$).

\subsection{An upper bound on the contribution of thermal GWs to dark radiation}\label{A:2}

We now evaluate the contribution to $\Delta N_{\rm eff}$ from thermal GWs, Eq~\eqref{eq:estY2}. Assuming a smooth transition to the radiation-dominated phase after the end of inflation, where both the bath temperature $T$ and Hubble parameter $H$ are monotonically decreasing functions of e-folds, we can estimate a conservative upper bound on $\Delta N_{\rm eff}$ by: (i) setting $[T(N_e^\rho)/T(N_e^{\rm RD}) \cdot e^{N_e^\rho - N_e^{\rm RD}}] = 1$ and (ii) evaluating it based on the values of $T$ and $H$ at the end of inflation rather than at the peak of the GW production $N_e^{\rm \rho}$, since both $T$ and $T/H$ will be greater at this earlier time. In formulas, we can write:
\begin{equation}
    \Delta N_{\rm eff}< \frac{4a_\nu}{g_\star}\Delta N_e^\rho\left[\frac{T(N_e^{\rm end})}{ H(N_e^{\rm end})}\right]\left[\frac{T(N_e^{\rm end})}{ \mpl}\right]^2, 
\end{equation}
where $N_e^{\rm end}$ corresponds to the end of inflation. Using the field value at the end of inflation from Eq~\eqref{eq:A4}, we can evaluate the corresponding values of the Hubble parameter and bath temperature by substituting into Eqs~\eqref{eq:A1} and \eqref{eq:A3}:
\begin{align}
    H(N_e^{\rm end}) &= \sqrt{\frac{\lambda_n}{3}}\left[\frac{n^2}{4(1+Q)}\right]^{\frac{n}{4}}\,M_{\rm pl}, \\
T(N_e^{\rm end}) &= \frac{30}{\pi^2g_\star}\left[\frac{\lambda_n Q}{4(1+Q)^2}\right]^{\frac{1}{4}}\left[\frac{n^2}{2^n(1+Q)}\right]^{\frac{n-2}{8}}\,M_{\rm pl}.
\end{align}

From these relations, we obtain the following conservative upper bound on $\Delta N_{\rm eff}$:
\begin{align}
    \Delta N_{\rm eff}&<\left[\frac{4a_\nu}{g_\star}\Delta N_e^\rho \left(\frac{30}{\pi^2g_\star}\right)^3\left(\frac{9\,n^{n-6}}{2^n}\right)^{\frac{1}{4}}\right]\nonumber \\
    &\times\frac{Q^{\frac{3}{4}}}{(1+Q)^{\frac{n+6}{8}}}\, \lambda_n^{\frac{1}{4}}. \label{eq:upperNeff2}
\end{align}
Here, the upper bound on $\Delta N_{\rm eff}$ depends on the height of the inflation potential, via $\lambda_n$, which is constrained by observational requirements. Specifically, we can determine the maximum allowed value of $\lambda_n$  using the results from Ref.~\cite{Montefalcone:2022owy}, which demonstrated that in the context of WI, the maximum allowed height of the inflaton potential required to avoid an overproduction of quantum fluctuations is always smaller than in standard cold inflation. In this sense, a conservative upper bound on $\lambda_n$ can be obtained using the corresponding cold inflationary bound which states:
\begin{equation}
    \left[\frac{H^2(N_e^{\rm cmb})}{|\dot{\phi}(N_e^{\rm cmb})|}\right]^2 \lesssim 4\pi^2A_s \label{eq:height_bound}
\end{equation}
where $A_s\simeq 2.1\times 10^{-9}$ corresponds to the amplitude of the power spectrum of scalar perturbations, obtained from CMB measurements~\cite{Planck:2018vyg}. Substituting the slow-roll solutions for $H$ and $\dot{\phi}$ at the CMB scale ($N_e^{\rm cmb}=60$ corresponding here to $N_e^{\rm ex}=0$) into Eq.~\eqref{eq:height_bound}, we obtain an upper bound on $\lambda_n$:
\begin{equation}
    \lambda_n\lesssim 2^{4+n}\pi^2A_s\frac{3(1+Q)^{\frac{n-2}{2}}}{n^n\left(1+4N_e^{\rm cmb}/n\right)^{\frac{2+n}{2}}}.
\end{equation}
Substituting this back into Eq.~\eqref{eq:upperNeff2} we obtain:
\begin{align}
\Delta N_{\rm eff}< C_n\,\frac{Q^{\frac{3}{4}}}{(1+Q)}A_s^{\frac{1}{4}},
\end{align}
with
\begin{align}
    C_n&\equiv \frac{8a_\nu}{g_\star}\Delta N_e^\rho \left(\frac{30}{\pi^2g_\star}\right)^3 \left(\frac{27\pi^2}{n^6}\right)^{\frac{1}{4}}\nonumber \\
    &\times \left(\frac{1}{1+4N_e^{\rm cmb}/n}\right)^{\frac{n+2}{8}},
\end{align}
which is a constant of order $\sim \mathcal{O}(10^{-7}-10^{-6})$ for $g_\star=106.75$.

We can rewrite the obtained bound on $\Delta N_{\rm eff}$ more explicitly, showing the dependence on $n$ and $Q$:
\begin{align}
    \Delta N_{\rm eff}&< 2.1\times 10^{-7} \left( \frac{106.75}{g_\star[T(N_e^{\rm ex})]}\right)^4\,\frac{Q^{\frac{3}{4}}}{1+Q} \nonumber \\
&\times \frac{1}{n^{\frac{3}{2}}\left(1+4\,N_e^{\rm cmb}/n\right)^{\frac{n+2}{8}}}.
\end{align}
This result is evidently very small, as the last two multiplicative factors are always less than unity for any $Q>0$ and $n \gtrsim \mathcal{O}(1)$. As a concrete example, setting $n=4$ and $N_e^{\rm cmb}=60$, and recalling that $Q^{3/4}/(1+Q)<1$, we get $\Delta N_{\rm eff}<  1.2\times 10^{-9}$. While this approximate analytical treatment does not exactly reproduce the results obtained from the full numerical evolution -- which can yield somewhat larger values, of the order $\sim\mathcal{O}(10^{-6})$ -- it clearly highlights the difficulty of producing a relevant contribution to dark radiation from thermal gravitons within single-field slow-roll WI scenarios.

\bibliographystyle{apsrev4-2}
\bibliography{reference}

\end{document}